%% file: S5R1_EAH.tex
\pdfoutput=1

\documentclass[aps,prd,showpacs,twocolumn,amssymb,amsmath,amsfonts,superscriptaddress,floatfix,preprintnumbers,altaffilletter, letterpaper]{revtex4}

\usepackage{graphicx}
\usepackage[raggedright,hang]{subfigure}
\usepackage{hyperref}
\usepackage{amssymb}
\usepackage{amsmath}
\usepackage{longtable}
\usepackage{rotating}
\usepackage{color}
\usepackage{epsfig}
\usepackage{epsf}
\usepackage{fancyhdr}
\usepackage{dcolumn}

\newcommand{\Tspanj}{T_{\mathrm{span},j}}

\newcommand{\Hz}{\mathrm{Hz}}
\newcommand{\second}{\mathrm{s}}
\newcommand{\muHz}{\mu\mathrm{Hz}}
\newcommand{\D}{\mathcal{D}}
\newcommand{\seg}{\mathrm{seg}}

\newcommand{\F}{\mathcal{F}}

\def\dcc{LIGO-P0900032-v3}
\DeclareGraphicsExtensions{.png}

\begin{document}

\pagestyle{fancy}

\rhead[]{}
\lhead[]{}

\title{
Einstein@Home search for periodic gravitational waves in early S5 LIGO data
}
 
\input{authorlist}

\date{\today}

\begin{abstract}
\noindent
This paper reports on an all-sky search for periodic gravitational waves
from sources such as deformed isolated rapidly-spinning neutron stars. 
The analysis uses $840$ hours of data from $66$ days of the fifth LIGO science run (S5).
The data was searched for quasi-monochromatic waves with frequencies~$f$ in the 
range from $50\,\Hz$ to $1500\,\Hz$, with a linear frequency drift $\dot{f}$ 
(measured at the solar system barycenter) in the range $-f/\tau < \dot f < 0.1\,f/\tau$, 
for a minimum spin-down age $\tau$ of $1\,000\,{\rm years}$ for signals below
$400\,\Hz$ and $8\,000\,{\rm years}$ above $400\,\Hz$.  The main
computational work of the search was distributed over approximately
$100\,000$ computers volunteered by the general public.  This large
computing power allowed the use of a relatively long coherent
integration time of $30\,{\rm hours}$ while searching a large
parameter space.  This search extends Einstein@Home's
previous search in LIGO S4 data
to about three times better sensitivity.
No statistically significant signals were found.  
In the $125\,\Hz$ to $225\,\Hz$ band, more than
$90\%$ of sources with dimensionless gravitational-wave strain tensor amplitude
greater than $3\times10^{-24}$ would have been detected.
\end{abstract}


\preprint{\dcc}

\maketitle

\section{Introduction\label{sec:introduction}}

Gravitational waves (GW) are predicted by Einstein's general
theory of relativity, 
but have so far eluded direct detection.
The Laser Interferometer Gravitational-wave Observatory (LIGO) 
\cite{ligo1,ligo2} has been built for this purpose and 
is currently the most sensitive gravitational-wave detector 
in operation.

Rapidly rotating neutron stars are expected to generate periodic 
gravitational-wave signals through various 
mechanisms~\cite{bildsten:1998,owen-1998-58,ushomirsky:2000,cutler:2002-66,jones-2002-331,melatos-2005-623,owen-2005-95}.
Irrespective of the emission mechanism, these signals are quasi-monochromatic 
with a slowly changing intrinsic frequency.
Additionally, at a terrestrial detector, such as LIGO, the data analysis problem
is complicated by the fact that the periodic GW signals are Doppler modulated by the 
detector's motion relative to the solar system barycenter~(SSB). 

A previous paper~\cite{S4EAH} reported on the results of the 
Einstein@Home search for periodic 
GW signals in the data from LIGO's fourth science run~(S4).
The present work extends this search, using more sensitive data
from $66$ days of LIGO's fifth science run~(S5).

Because of the weakness of the GW signals buried in the detector noise,
the data analysis strategy is critical. A powerful detection method is given by
coherent matched-filtering. This means one convolves all available data
with a set of template waveforms corresponding to all possible putative sources.
The resulting detection statistic is derived in Ref.~\cite{jks} and is commonly 
referred to as the $\F$-statistic.

The parameter space to be scanned for putative signals from 
isolated neutron stars is four-dimensional,
with two parameters required to describe the source sky position using 
standard astronomical equatorial coordinates $\alpha$ (right ascension) 
and $\delta$ (declination), and additional coordinates
$(f, \dot{f})$ denoting the intrinsic frequency and frequency drift.
To achieve the maximum
possible sensitivity, the template waveforms must match the source
waveforms to within a fraction of a cycle over the entire observation
time (months or years for current data samples). 
So one must choose a very closely spaced grid of templates in this
four-dimensional parameter space. This makes the computational cost
of the search very high, and therefore limits the search sensitivity~\cite{jk3}.  

To maximize the possible integration time, and hence achieve 
a more sensitive search, the computation was distributed via
the volunteer computing project Einstein@Home~\cite{EaHURL}.
This large computing power allowed the use of a relatively long coherent
integration time of $30\,{\rm h}$, despite the large parameter space searched.  
Thus, this search involves coherent matched-filtering in the form of the $\F$-statistic over 
$30$-hour-long data segments and subsequent incoherent combination of $\F$-statistic results
via a coincidence strategy.

The methods used here are further described in 
Sections~\ref{sec:DataPrep}-\ref{sec:PostProc}.
Estimates of the sensitivity of this search and results are in Sections~\ref{sec:ExpectSen}
and~\ref{sec:Results}, respectively. Previously, other all-sky searches for periodic
GW sources using LIGO S4 and S5 data, which combine power from many
short coherent segments (30-minute intervals) of data, have been
reported by the LIGO Scientific Collaboration (LSC) \cite{pshS4:2008,PFearlyS5}.  
However, this Einstein@Home search explores large regions of parameter space
which have not been analyzed previously with LIGO S5 data. 
The sensitivity of the results here are compared with previous searches
in Section~\ref{sec:Comparison}, and conclusions are given in Section~\ref{sec:Conclusion}.

\section{Data selection and preparation
\label{sec:DataPrep}}

The data analyzed in the present work was collected
between November 19, 2005 and January 24, 2006.
The total data set covering frequencies from $50\,\Hz$
to $1500\,\Hz$ consisted of $660\,{\rm h}$ of data from
the LIGO Hanford \mbox{4-km~(H1) detector} and $180\,{\rm h}$
of data from the LIGO Livingston \mbox{4-km~(L1) detector}.

The data preparation method is essentially identical to that of the previous
S4 analysis~\cite{S4EAH}. Therefore only a  brief summary of the main
aspects is given here;
further details are found in~\cite{S4EAH} and references therein.
The data set has been divided into segments of~$30\,{\rm h}$ each. 
 However, the 30-hour long data
segments are not contiguous, but have time gaps.
Since the number
of templates required increases rapidly with observation span,
the $30\,{\rm h}$ of data for each segment were chosen to lie within
a time span of less than $40\,{\rm h}$. In what follows, the notion
of  ``segment" will always refer to one of these time stretches,
each of which contains exactly $T=30\,{\rm h}$ of data. The total
time spanned by a given data segment~$j$ is denoted by~$\Tspanj$ and
conforms to $30\,{\rm h} < \Tspanj< 40\,{\rm h}$.

Given the above constraints, a total of $N_\seg=28$ data segments
($22$ from H1, $6$ from L1) were obtained from the early S5 data considered.
These data segments are labeled by $j=1,\dots,28$. Table~\ref{t:datasegs}
lists the global positioning system~(GPS) start time along with
the time span of each segment.

\begin{table}
\caption{ \label{t:datasegs} Segments of early S5 data
  used in this search. The
  columns are the data segment index $j$, the GPS start time $t_j$
  and the time spanned~$\Tspanj$.
  }
\begin{center}
\begin{tabular}{ccccc}
\hline
$j$ & Detector  & $t_j$ [s] & $\Tspanj$ [s] \\
\hline\hline
1 & H1 &816397490 & 140768\\
2 & H1 &816778879 & 134673\\
3 & H1 &816993218 & 134697\\
4 & H1 &817127915 & 137962\\
5 & H1 &817768509 & 142787\\
6 & H1 &817945327 & 143919\\
7 & H1 &818099543 & 139065\\
8 & H1 &818270501 & 143089\\
9 & H1 &818552200 & 134771\\
10 & H1 &818721347 & 138570\\
11 & H1 &818864047 & 134946\\
12 & H1 &819337064 & 143091\\
13 & H1 &819486815 & 120881\\
14 & H1 &819607696 & 116289\\
15 & H1 &819758149 & 136042\\
16 & H1 &820482173 & 143904\\
17 & H1 &820628379 & 138987\\
18 & H1 &821214511 & 126307\\
19 & H1 &821340818 & 126498\\
20 & H1 &821630884 & 141913\\
21 & H1 &821835537 & 138167\\
22 & H1 &821973704 & 142510\\
23 & L1 &818812286 & 130319\\
24 & L1 &819253562 & 140214\\
25 & L1 &819393776 & 126075\\
26 & L1 &819547883 & 138334\\
27 & L1 &820015400 & 121609\\
28 & L1 &821291797 & 140758\\
\hline
\end{tabular}
\end{center}
\end{table}

In this analysis, the maximum frequency shift of a signal over
the length of any given data segment and parameter-space range
examined is dominated by the Doppler modulation due
to the Earth's orbital motion around the solar system barycenter (SSB),
while the effects of frequency change resulting from intrinsic spin-down
of the source are smaller. The orbital velocity of the Earth
is about $v/c\approx 10^{-4}$, hence a signal will always remain
in a narrow frequency band smaller than $\pm0.15\,\Hz$ around
a given source frequency. Therefore, for each detector the total 
frequency range from $50\,\Hz$ to $1500\,\Hz$ is broken up into 2900 slices, 
each of $0.5\,\Hz$ bandwidth plus overlapping wings of $0.175\,\Hz$ on either side. 

\begin{figure}
	\includegraphics[width=1.0\columnwidth]{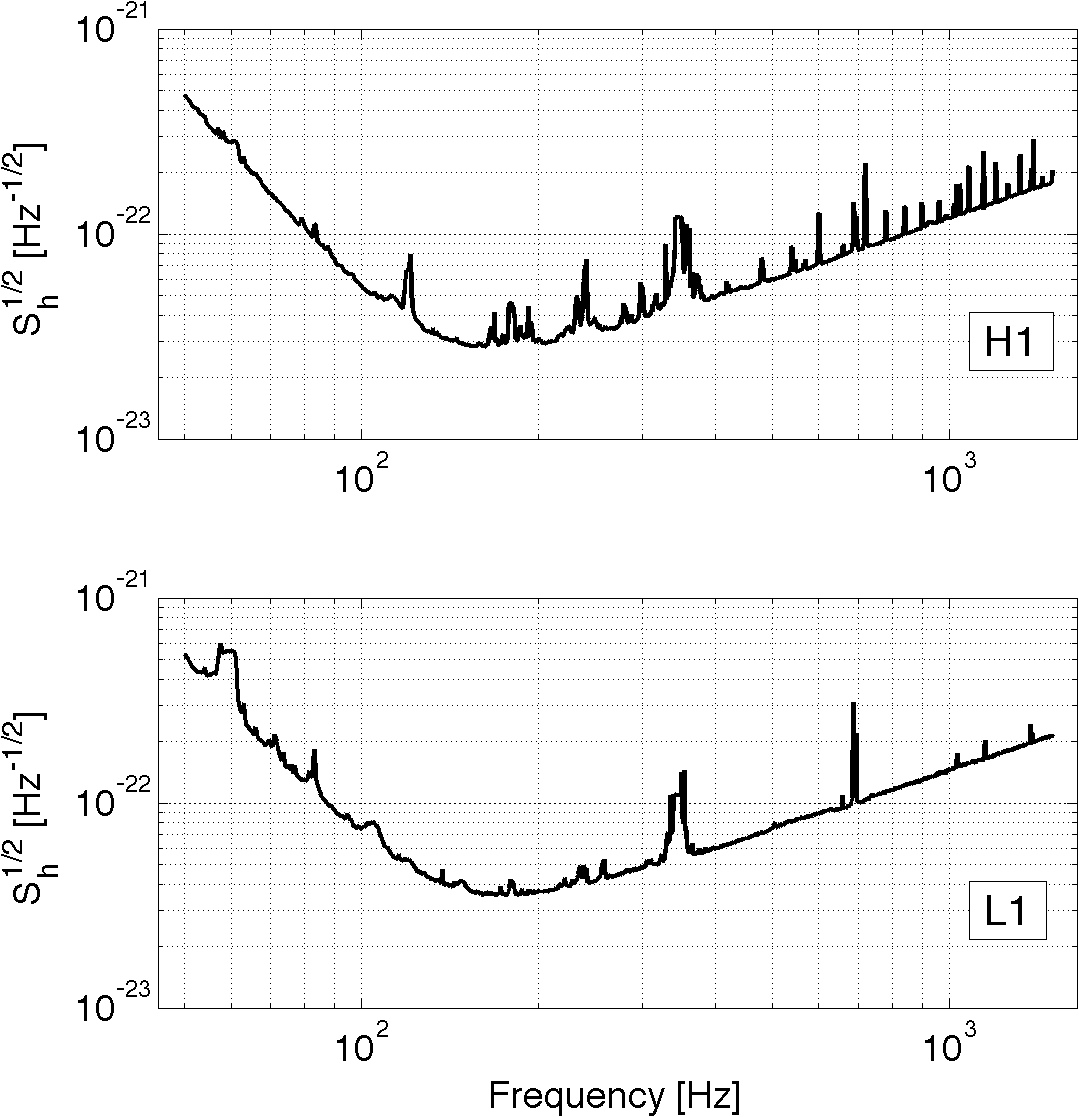}
	\caption{Strain amplitude spectral densities $\sqrt{S_{\rm h}(f)}$ of 
	the cleaned data from the LIGO detectors H1 (top)
         and L1 (bottom) used in the Einstein@Home searches. 
         The curves in the top (bottom) panel are the harmonic mean of 
         the 22 H1 (6 L1) 30-hour segments of S5 data used this Einstein@Home analysis. 
         }
	\label{f:noisefloors}
\end{figure}

The detector data contains numerous narrow-band noise artifacts,
so-called ``lines", which are of instrumental origin, such as
harmonics of the $60\,\Hz$ mains frequency. Prior to the analysis,
line features of understood origin (at the time before the launch of the search) 
were removed (``cleaned") from the data by substitution of the 
frequency-domain data bins with random
Gaussian noise. Table~\ref{t:lines} in the Appendix shows the frequencies of
lines excluded from the data. The harmonic mean noise strain amplitude spectra of the final
cleaned H1 and L1 data sets are shown in Fig.~\ref{f:noisefloors}.

\section{Data processing
\label{sec:DataProc}}

The paper describing the previous Einstein@Home search in S4 
data~\cite{S4EAH} presented in detail the data processing scheme. For the purpose of
the present search the same data processing infrastructure is employed.
Hence, here only a short summary thereof is given, pointing out the minimal
changes applied in setting up the present analysis.

The total computation of the search is broken up into $16\,446\,454$ workunits.
Each workunit represents a separate computing task and is processed using
the Berkeley Open Infrastructure for Network Computing
(BOINC)~\cite{BOINC,BOINC1,BOINC2}. To eliminate errors and weed out
results that are wrong,
each workunit
is independently processed by at least two different volunteers.
Once two successful results for a workunit are returned back to the 
Einstein@Home server, they are compared by an automatic validator,
which discards results that differ by more than some allowed tolerance.
New workunits are generated and run independently again for such cases.

In searching for periodic gravitational-wave signals, each workunit examines a different part of 
parameter space. A key design goal is that the computational effort to
conduct the entire analysis should take about 6--7 months. An additional
design goal is to minimize the download burden on the Einstein@Home 
volunteers' internet connections and also on the Einstein@Home data
servers. This is accomplished by letting each workunit  use 
only a small re-usable subset of the total data set, so that
Einstein@Home volunteers are able to carry out useful
computations on a one-day time scale.

Each workunit searches only one data segment over a narrow frequency range,
but covering all of the sky and the entire range of frequency derivatives.
The workunits are labeled
by three indices~$(j,k,\ell)$, where $j=1,\dots,28$ denotes the data segment,
$k=1,\dots,2900$ labels the $0.5\,\Hz$ frequency band and $\ell=1,\dots,M(j,k)$
enumerates the individual workunits pertinent to data segment~$j$ and frequency
band~$k$.

In each segment the $\F$-statistic is evaluated on a grid in parameter space. 
Each parameter-space grid is constructed such that grid points (templates) are not further apart 
from their nearest neighbor by more than a certain distance. The distance measure is defined from 
a metric on parameter space, first introduced in \cite{metric1,metric2},
representing the fractional loss of squared signal-to-noise ratio (${\rm SNR}^2$) due to
waveform mismatch between the putative signal and the template.
For any given workunit, the parameter-space grid is a
Cartesian product of uniformly-spaced steps $d f$ in frequency,
uniformly-spaced steps $d \dot{f}$ in frequency derivative,
and a two-dimensional  sky grid, which has non-uniform spacings
determined by the metric~\cite{prix:eahdoc,S4EAH}.

For frequencies in the range
$[50,400)\,\Hz$, the maximal allowed mismatch was chosen as $m = 0.15$
(corresponding to a maximal loss in ${\rm SNR}^2$ of $15\%$), while in the range
$[400,1500)\,\Hz$, the maximal mismatch was $m = 0.4$. 
It can be shown~\cite{prix:eahdoc,S4EAH}, that these choices of maximal mismatch 
enable a coherent search of near-optimal sensitivity at fixed computational resources.

The step-size in frequency~$f$ obtained from the metric depends
on $\Tspanj$ of the $j$'th data segment: \mbox{$d f_j =2\sqrt{3m}/(\pi\Tspanj)$}.
In the low-frequency range this results in frequency spacings in the
range \mbox{$d f_j \in [ 2.97,\,3.67]\,\muHz$}, 
while for high-frequency workunits \mbox{$d f_j \in [4.85,\,6.0]\,\muHz$}.

The range of frequency derivatives $\dot f$ searched is defined in
terms of the ``spin-down age'' \mbox{$\tau \equiv - f/\dot f$}, namely
\mbox{$\tau \ge 1000\,$years} for low-frequency and \mbox{$\tau \ge 8\,000$\,years}
for high-frequency workunits.
As in the S4 Einstein@Home search, these ranges were guided by the
assumption that a nearby very young neutron star would correspond to a
historical supernova, supernova remnant, known pulsar, or pulsar wind nebula.
The search also covers a small ``spin-up'' range, so the actual ranges
searched are \mbox{$\dot{f} \in [-f/\tau,\,0.1 f/\tau]$}.
In $\dot f$ the grid points are spaced according to
\mbox{$d\dot{f}_j = 12 \sqrt{5m}/(\pi\,\Tspanj^2)$},
resulting in resolutions $d\dot{f}_j \in [1.60,\,2.44]\times 10^{-10}\,\Hz/\second$
for low-frequency workunits, and
\mbox{$d\dot{f}_j \in [2.61,\, 3.99]\times 10^{-10}\,\Hz/\second$} for high-frequency
workunits.

The resolution of the search grid in the sky
depends both on the start time $t_j$ and
duration $\Tspanj$ of the segment, as well as on the frequency~$f$. The
number of grid points on the sky scales as~$\propto f^2$,
and approximately as~$\propto \Tspanj^{2.4}$
for the range of \mbox{$\Tspanj \sim 30-40\,{\rm h}$} used in this search.
As was done in the previous S4 analysis~\cite{S4EAH}, to simplify the 
construction of workunits and
limit the number of different input files to be sent, the  sky grids are fixed
over a frequency range of $10\,\Hz$, but differ for each data
segment~$j$. The  sky grids are computed at the higher end of each
10-Hz band, so they are slightly ``over-covering'' the sky at lower
frequencies within the band. The search covers in total a frequency band of
$1450\,\Hz$, so there are 145 different  sky grids for each segment.

The output from one workunit in the low (high) frequency 
range contains the top $1\,000$ ($10\,000$) candidate events with the
largest values of the $\F$-statistic. In order to balance the load
on the Einstein@Home servers, a low-frequency workunit
returns a factor of $10$ fewer events, because low-frequency
workunits require runtimes
approximately $10$ times shorter than high-frequency workunits.
For each candidate event
five values are reported: frequency (Hz), right ascension angle (radians), 
declination angle (radians), frequency derivative (Hz/s) and 
$2\F$ (dimensionless). The frequency is the frequency at the SSB at the
instant of the first data point in the corresponding data segment.
Returning only the ``loudest'' candidate events effectively
corresponds to a floating threshold on the value of the
$\F$-statistic.  This avoids large lists of candidate events being
produced in regions of parameter space containing non-Gaussian noise,
such as instrumental artifacts that were not removed {\it a priori}
from the input data because of unknown origin.

\section{Post-processing
\label{sec:PostProc}}

After results for each workunit are returned to the Einstein@Home servers 
by project volunteers, post-processing is conducted on those servers 
and on dedicated computing clusters.  The  post-processing has the goal
of finding candidate events that appear in
many of the $28$ different data segments with consistent parameters.

In this search, the post-processing methods are the same as used
for the Einstein@Home S4 search~\cite{S4EAH}.
Therefore, this section only summarizes the main steps;
a more detailed description can be found in~\cite{S4EAH}.

A consistent (coincident) set of ``candidate events''
is called a ``candidate''.  Candidate events from different data
segments are considered coincident if they cluster closely together in
the four-dimensional parameter space.  
By using a grid of ``coincidence cells'', the clustering method 
can reliably detect strong  signals, which would produce candidate 
 events with closely-matched parameters in many of the $28$ data segments.
The post-processing pipeline operates in $0.5\,\Hz$-wide 
frequency bands, and performs the following steps described below.

\subsection{The post-processing steps}

A putative source with non-zero spin-down would generate
candidate events with different apparent frequency values
in each data segment. To account for these effects, the
frequencies of the candidate events are shifted back to the
same frequency value at fiducial time $t_{\rm{fiducial}}$ via 
\mbox{$f(t_{\rm{fiducial}}) =  f(t_{j}) +  (t_{\rm{fiducial}} - t_{j}) \, \dot{f}$},
where $\dot{f}$ and $f(t_j)$ are the spin-down rate and frequency of a
candidate event reported by the search code in the result file, and
$t_j$ is the time-stamp of the first datum in the $j$'th~data segment.  
 The fiducial time is chosen to be the GPS start time
of the earliest ($j=1$) data segment, 
\mbox{$t_{\rm{fiducial}} = t_1 = 816\,397\,490\,{\rm s}$}.

A grid of cells is then constructed  in the
four-dimensional parameter space to find coincidences among the
28 different data segments. The coincidence search algorithm uses 
rectangular cells in the coordinates $(f, \dot f, \alpha \cos \delta, \delta)$.
The dimensions of the cells are adapted to the parameter-space search grid
(see below).
Each candidate event is assigned to a
particular cell.  In cases where two or more candidate events from the
same data segment $j$ fall into the same cell, only the candidate
event having the largest value of $2\F$ is retained in the cell.  Then
the number of candidate events per cell coming from distinct data
segments is counted, to identify cells with more coincidences than
would be expected by random chance.

To ensure that candidate events located on opposite sides
of a cell border are not missed, the entire cell coincidence grid is
shifted by half a cell width in all possible $2^4 = 16$~combinations
of the four parameter-space dimensions.  Hence, $16$ different
coincidence cell grids are used in the analysis.

\subsection{Construction of coincidence windows}

The coincidence  cells are constructed to be as small as
possible to reduce the probability of false alarms.  
However, since each of the $28$ different data segments uses a
different parameter space grid, the coincidence cells must be chosen
to be large enough that the candidate events from a source (which
would appear at slightly different points in parameter space in each
of the $28$ data segments) would still lie in the same coincidence cell.

In the frequency direction, the size $\Delta f$ for the coincidence cell is
given by the largest search grid spacing in~$f$ (for smallest value of $\Tspanj$) 
plus the largest possible offset in spin-down:
\mbox{$\Delta f = \; \max_{j} \, ( df_j +  \Delta t \; d\dot f_j )$},
where the maximization over $j$ selects the data segment with the
smallest $T_{{\rm span},j}$ (which is $j=6$) and
\mbox{$\Delta t = |\max_{j} \, t_{j}  - \min_{j} \, t_{j} | = t_{22} - t_{1}  \;  = 5\,576\,214 \, {\rm s}$}
is the total time span between the latest and earliest data segments.
For safety, e.g.\ against noise fluctuations that could shift a candidate peak, 
$\Delta f$ has been increased by a further 30\%, so that the width of the 
coincidence cell in $f$ below $400\,\Hz$ is $\Delta f = 1.78\,{\rm mHz}$ 
and $\Delta f = 2.9\,{\rm mHz}$ above $400\,\Hz$.

In the frequency-derivative direction, the size of the coincidence cell
is given by the largest $d\dot f_j$ spacing in the parameter space
grid, which is also determined by the smallest value of $\Tspanj$.  
For safety this is also increased by $30\%$, so that 
$\Delta\dot f = 3.18 \times 10^{-10} \,\textrm{ Hz  s}^{-1} $ below
$400\,\Hz$ and 
$\Delta\dot f = 5.19 \times 10^{-10}\, \textrm{ Hz s}^{-1}$ above $400\,\Hz$ .

\begin{figure}
	\includegraphics[width=1.0\columnwidth]{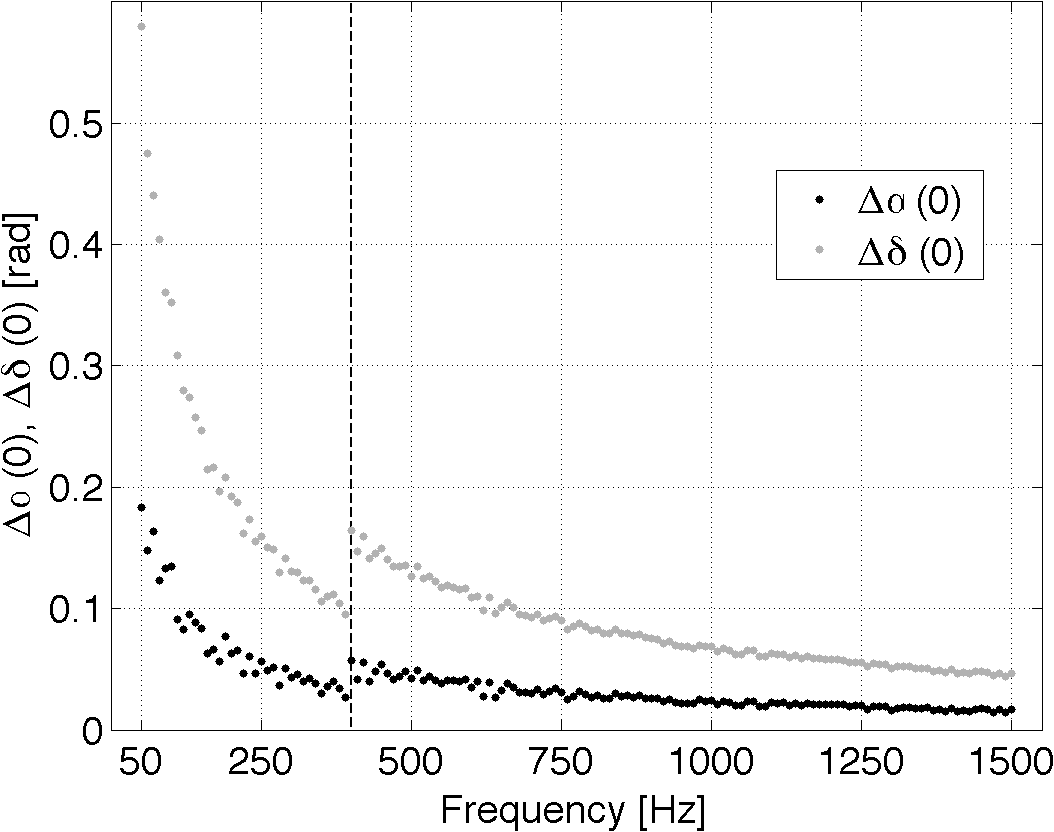}
	\caption{ The parameters $\Delta \alpha(0)$ and $\Delta
          \delta(0)$ of the sky coincidence-window model
          as a function of the 10~Hz frequency band. The vertical
          dashed line at $400\,\Hz$ indicates the separation between the low and high
          frequency ranges.}
\label{f:SkyCoinParams}
\end{figure}

In sky position, the size of the coincidence cells is guided by the
behavior of the parameter-space metric. As described in~\cite{S4EAH},
the density of grid points in the sky is approximately proportional
to \mbox{$|\cos(\delta)\,\sin(\delta)| \propto |\sin(2\delta)|$}, and it follows
from~\cite{S4EAH} that \mbox{$\cos(\delta) \,d\alpha = |\sin(\delta)| \,d\delta = {\rm const}$}.
Because of the singularity when $\delta \to 0$, a useful model
for the coincidence window size varying with declination
is given by
\begin{eqnarray} 
\Delta \alpha(\delta) & = &\Delta \alpha(0) / \cos (\delta)\label{e:alphawin} \\ 
\Delta \delta (\delta)& = &\Biggl\{
\begin{array}{lc}
  \Delta \delta(0)& \mbox{ if } |\delta| < \delta_{\rm c},\\
  \Delta \alpha(0) / |\sin(|\delta| - \kappa\,  \Delta \alpha(0))| &\mbox{ if }  |\delta| \ge \delta_{\rm c}.
\end{array} \nonumber
\end{eqnarray}
To ensure continuity at $\delta = \delta_{\rm c}$, the transition
point $\delta_{\rm c}$ is defined by the condition 
$\Delta \alpha(0) / |\sin(|\delta_c| - \kappa\, \Delta \alpha(0))| = \Delta \delta(0)$.
The  tuning parameter~$\kappa$ is chosen based on visual inspection
to be $\kappa=1.5$ in this search.
The values of $\Delta \alpha(0)$ and $\Delta \delta(0)$ are directly
determined from the sky grids (see \cite{S4EAH} for details).
Figure~\ref{f:SkyCoinParams} shows these parameters for all sky grids 
as a function of frequency. As stated above, 
the sky grids are constant for $10\,\Hz$-wide steps in
frequency, and so these parameters vary with the same step-size.

\subsection{Output of the post-processing}

The output of the post-processing is a list of the candidates with the
greatest number of coincidences.
The possible number of coincidences ranges from a 
minimum of $0$ to a maximum of $28$ (the number of data segments analyzed). 
The meaning of~$\mathcal{C}$ coincidences is that there are
$\mathcal{C}$~candidate events within a given coincidence cell.
In each frequency band of coincidence-window
width $\Delta f$, the coincidence cell containing the largest number
of candidate events is found.  
The pipeline outputs the
average frequency of the coincidence cell, the average sky position
and spin-down of the candidate events, the number of candidate events
in the coincidence cell, and the ``significance'' of the candidate.
The significance of a candidate, first introduced
in~\cite{S2FstatPaper} and explained in~\cite{S4EAH}, is defined by
\begin{equation}
 \mathcal{S} = \sum_{q=1}^{\mathcal{C}} (\F_q - \ln(1+\F_q))
 \label{e:significance} \; ,
\end{equation}
where $\F_q$ is the $\F$-statistic value of the $q$'th candidate event
in the same coincidence cell, which harbors a total of 
$\mathcal{C}$ candidate events.

 \subsection{False alarm probability and detection threshold}
 
The central goal of this search is to make a {\it confident detection}, not to
set upper limits with the broadest possible coverage band.  This is
reflected in the choice of detection threshold
based on the expected false alarm rates.
In this search the background level of false alarm candidates is expected at 
$10$ coincidences (out of $28$ possible). 
As a pragmatic choice,  
the threshold of confident detection is set at
$20$ coincidences, which is highly improbable to arise from random noise only.
These settings will be elucidated in the following.
 
To calculate the false alarm probabilities, consider the case where
$\mathcal{E}_{\rm seg}(k)$ candidate events per data segment 
obtained from pure Gaussian noise are 
distributed uniformly about $N_{\rm cell}(k)$ independent coincidence cells
in a given $0.5\,\Hz$ band~$k$. 
Assuming the candidate events
are independent, the probability $p_{\rm F}(k;\mathcal{C}_{\rm max})$ per coincidence cell of finding
$\mathcal{C}_{\rm max}$ or more candidate events from different data segments has been
derived in~\cite{S4EAH} and is given by the binomial distribution
\begin{equation} 
 p_{\rm F}(k;\mathcal{C}_{\rm max}) = \sum_{n=\mathcal{C}_{\rm max}}^{N_{\rm seg}} {N_{\rm seg} \choose n} 
 [\epsilon(k)]^n [1-\epsilon(k)]^{N_{\rm seg} - n}\,,
\end{equation}
where $\epsilon(k)$ denotes the probability of populating any given coincidence cell
with one or more candidate events in a given data segment, obtained as
\begin{equation}
  \epsilon(k) = 1 - \biggl(1-{1 \over N_{\rm cell}(k)} \biggr)^{\mathcal{E}_{\rm seg}(k)}\,.
\end{equation}
Finally, the probability~$P_{\rm F}(k;\mathcal{C}_{\rm max})$ that there are~$\mathcal{C}_{\rm max}$
or more coincidences in {\em one or more} of the $N_{\rm cell}$ cells per $0.5\,\Hz$ band $k$ is 
\begin{equation}
  P_{\rm F}(k;\mathcal{C}_{\rm max}) = 1-\left[1-p_{\rm F}(k;\mathcal{C}_{\rm max})\right]^{N_{\rm cell}}.
  \label{e:FArate}
\end{equation}

\begin{figure}
	\includegraphics[width=0.98\columnwidth]{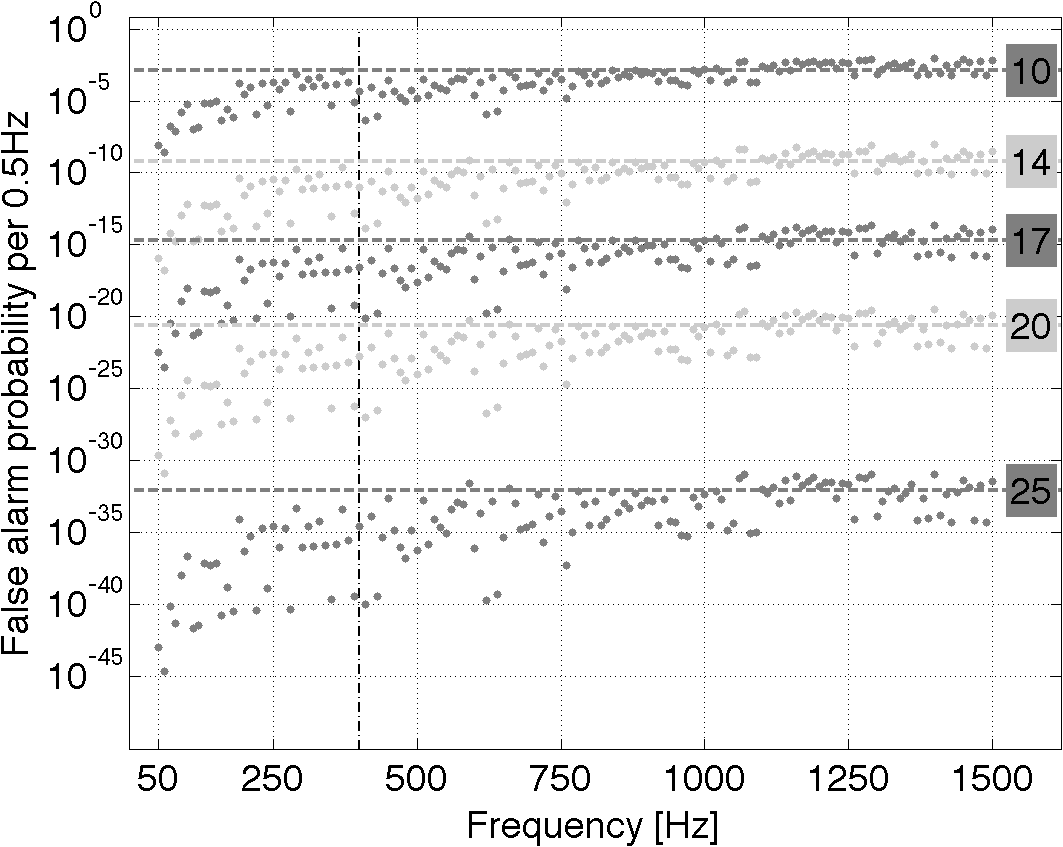}
	\caption{ False alarm probabilities~$P_{\rm F}(k;\mathcal{C}_{\rm max})$ as a function
	 of frequency band (labeled by $k$) for different values of 
	 \mbox{$\mathcal{C}_{\rm max}\in\{10,14,17,20,25\}$}. 
	 The dashed horizontal lines represent the corresponding average across all
	 frequencies.
	 The vertical dashed line at $400\,\Hz$ indicates the separation 
	 between the low and high frequency ranges.
	\label{f:FA}}
\end{figure}

Figure~\ref{f:FA} shows the dependence of $P_{\rm F}(k;\mathcal{C}_{\rm max})$ on the
frequency bands for different values of~$\mathcal{C}_{\rm max}$.
One finds that the average false alarm probability of 
obtaining $10$ or more
coincidences is approximately $10^{-3}$. This means,
in our analysis of $2\,900$ half-Hz frequency bands,
only a few candidates are expected to have $10$ or more coincidences.
Thus this will be the anticipated background level 
of coincidences, because from pure random noise one would not expect
candidates of {\em more than} $10$ coincidences in this analysis.
In contrast, the false alarm probability of reaching the detection threshold
of $20$ or more coincidences per $0.5\,\Hz$ averaged over all 
frequency bands is about~$10^{-21}$.
Therefore, this choice of detection threshold makes it extremely improbable
to be exceeded in case of random noise.

\section{Estimated Sensitivity\label{sec:ExpectSen}}

The methods used here would be expected to yield very high confidence if
a strong signal were present.  To estimate
the sensitivity of this detection scheme,
Monte-Carlo methods are used to simulate a population of sources.  
The goal is to find the strain
amplitude $h_0$ at which $10\%$, $50\%$, or $90\%$ of sources uniformly
populated over the sky and in their ``nuisance parameters'' 
would be confidently detected.
In this analysis,
``detectable'' means ``produces coincident events in $20$ or more
distinct data segments''.
As discussed above, the false alarm probability for obtaining
such a candidate in a given $0.5\,\Hz$ band is of order $10^{-21}$.
This is therefore an
estimate of the signal strength required for high-confidence
detection. For this purpose, the pipeline developed 
in~\cite{S4EAH} is run here, using the input data of the present analysis.
A large number of distinct simulated sources (trials) are tested
for detection. A ``trial" denotes a single simulated source
which is probed for detection. For a detailed description of 
the methodology, the reader is referred to~\cite{S4EAH}.

\begin{figure}
   \includegraphics[width=1.0\columnwidth]{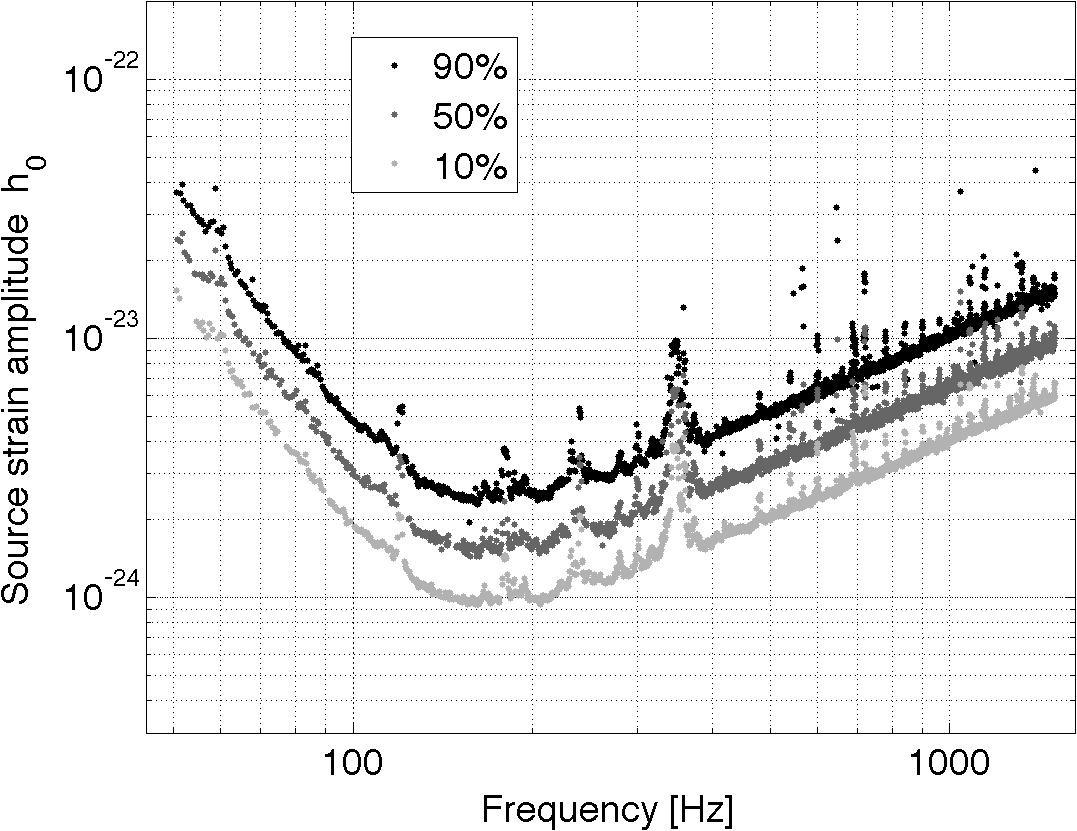}
   \caption{Estimated sensitivity of the Einstein@Home search for isolated periodic GW
     sources in the early LIGO S5 data.  The set of three curves shows
     the source strain amplitudes $h_0$ at which $10\%$ (bottom), $50\%$
     (middle) and $90\%$ (top) of simulated sources would be confidently
     detected in this Einstein@Home search.}
   \label{f:estULs}
\end{figure}

Figure~\ref{f:estULs} shows the resulting search sensitivity
curves as functions of frequency. Each data point on the plot denotes
the results of $1\,000$ independent trials.  These show the values of
$h_0$ as defined in~\cite{jks} such that $10\%$, $50\%$, and $90\%$ of
simulated sources are confidently detected in the post-processing pipeline.

The dominant sources of error in these sensitivity curves are uncertainties 
in calibration of the LIGO detector response functions (cf.~\cite{S4EAH,PFearlyS5}). 
The uncertainties range typically from about $8\%$ to $15\%$, depending on frequency.

The behavior of the curves shown in Fig.~\ref{f:estULs} essentially reflects the
instrument noise given in Fig.~\ref{f:noisefloors}. 
One may fit the curves obtained in Fig.~\ref{f:estULs} to the shape of the 
harmonic-mean averaged strain noise power spectral density $S_h(f)$.  
Then the three sensitivity curves in Fig.~\ref{f:estULs} 
are described by
\begin{equation}
  h_0^{\D} (f) \approx R_{\D} \; \sqrt{\frac{S_h(f)}{30\; {\rm h}}}\,,
  \label{e:sensitivity}
\end{equation}
where the pre-factors $R_{\D}$ for different detection probabilities
levels $\D=90\%$, $50\%$, and $10\%$ are well fit below $400\,\Hz$ by
$R_{90\%} = 29.4$, $R_{50\%} = 18.5$, and $R_{10\%} = 11.6$, 
and above $400\,\Hz$ by 
$R_{90\%} = 30.3$, $R_{50\%} = 19.0$, and \mbox{$R_{10\%} = 11.8$}.

\section{Results \label{sec:Results}}

\subsection{Vetoing instrumental noise lines\label{ssec:Veto}}
At the time the instrument data was prepared and cleaned, narrow-band
instrumental line features of known origin were removed, as previously
described in Sec.~\ref{sec:DataPrep}.  However, the data also
contained stationary instrumental line features that were not
understood, or were poorly understood, and thus were not removed 
{\it a priori}.  After the search had been conducted,
at the time the post-processing started, the origin of
more stationary noise lines became known. Therefore, these
lines, whose origin was tracked down after the search, are
excluded (cleaned {\it a posteriori}) from the results.
A list of the polluted frequency bands which have been
cleaned {\it a posteriori} is shown in Tab.~\ref{t:postcleaning}
in the Appendix.

However, noise features still not understood instrumentally at this
point were not removed from the results.  As a consequence, the
output from the post-processing pipeline contains instrumental
artifacts that in some respects mimic periodic GW signals.  But these artifacts
tend to cluster in certain regions of parameter space, and in many
 cases they can be automatically identified and vetoed as done
in previous searches~\cite{S4IncoherentPaper,S4EAH}.
The method used here is derived in~\cite{PletschGC} and
a detailed description of its application is found in~\cite{S4EAH}.

For a coherent observation time baseline of $30\,{\rm h}$ the parameter-space
regions where instrumental lines tend to appear are determined by
global-correlation hypersurfaces~\cite{PletschGC} of the $\F$-statistic.
On physical grounds, in these parameter-space regions there
is little or no frequency Doppler modulation from the Earth's motion,
which can lead to a relatively stationary detected frequency.
Thus, the locations of instrumental-noise candidate events are described by
\begin{equation}
  \left| \,\dot f + f\;\frac{\boldsymbol{v}_j} {c} \cdot {\hat{\boldsymbol{n}}} \,\right| < \epsilon \;,
  \label{e:vetocond}
\end{equation}
where $c$ denotes the speed of light, $  {\hat{\boldsymbol{n}}}$ is a unit vector pointing
to the source's sky-location in the SSB frame and 
relates to the equatorial coordinates~$\alpha$ and~$\delta$ by  
${\hat{\boldsymbol{n}}} = (\cos \delta \, \cos \alpha, \cos \delta \, \sin \alpha, \sin \delta)$, 
$\boldsymbol{v}_j$ is the orbital velocity of the
Earth at the midpoint of the $j$'th data segment ($| \boldsymbol{v}_j| \approx 10^{-4} \,c$).
The parameter $\epsilon$ accounts for a certain tolerance
needed due to the parameter-space gridding and
can be understood as \mbox{$\epsilon =\Delta f/N_{\rm c}\,\Delta T$},
where $\Delta f$ denotes width in frequency (corresponding
to the coincidence-cell width in the post-processing) up to
which candidate events can be resolved during the 
characteristic length of time $\Delta T$, and $N_{\rm c}$
represents the size of the vetoed or rejected region, measured in coincidence cells.
In this analysis $\Delta T = 5\,718\,724\, {\rm s}$ 
($\approx 66\,{\rm days}$) is the total time interval spanned by the
input data.

Because false alarms are expected at
the level of $10$ coincidences, candidates that satisfy Eq.~(\ref{e:vetocond})
for more than $10$ data segments are eliminated (vetoed).
The fraction of parameter space excluded by this veto
is determined by Monte-Carlo simulations to be
about $13\%$. From Eq.~(\ref{e:vetocond}) it follows that for fixed frequency 
the resulting fraction of sky excluded by the veto (uniformly averaged over spin-down)
is greatest at lowest frequencies and decreases approximately as $f^{-1}$ for higher frequencies.
Appendix~A of Ref.~\cite{S4EAH} presents an example calculation, illustrating
the parameter-space volume excluded by this vetoing method.

\subsection{Hardware-injected signals \label{ssec:HwInj}}

During parts of the LIGO S5 run ten simulated periodic GW signals
were injected at the hardware level, by modulating the
interferometer mirror positions via signals sent to 
voice actuation coils surrounding magnets glued near the
mirror edges.
The hardware injections were active only part of the time
during the S5 science run; 
in only $12$ (of the $28$) 
data segments chosen for this search did the hardware injections have  
duty cycles greater than $90\%$.
But the value of $12$ coincidences is far below the detection condition. 
Therefore, the hardware
injections are not expected to be detected in this search, 
simply because they were
inactive during a large fraction of the data analyzed.

\subsection{Post-processing results \label{ssec:PpResults}}

\begin{figure}
	\subfigure[\label{f:skym-a}]
	{\includegraphics[width=0.85\columnwidth]{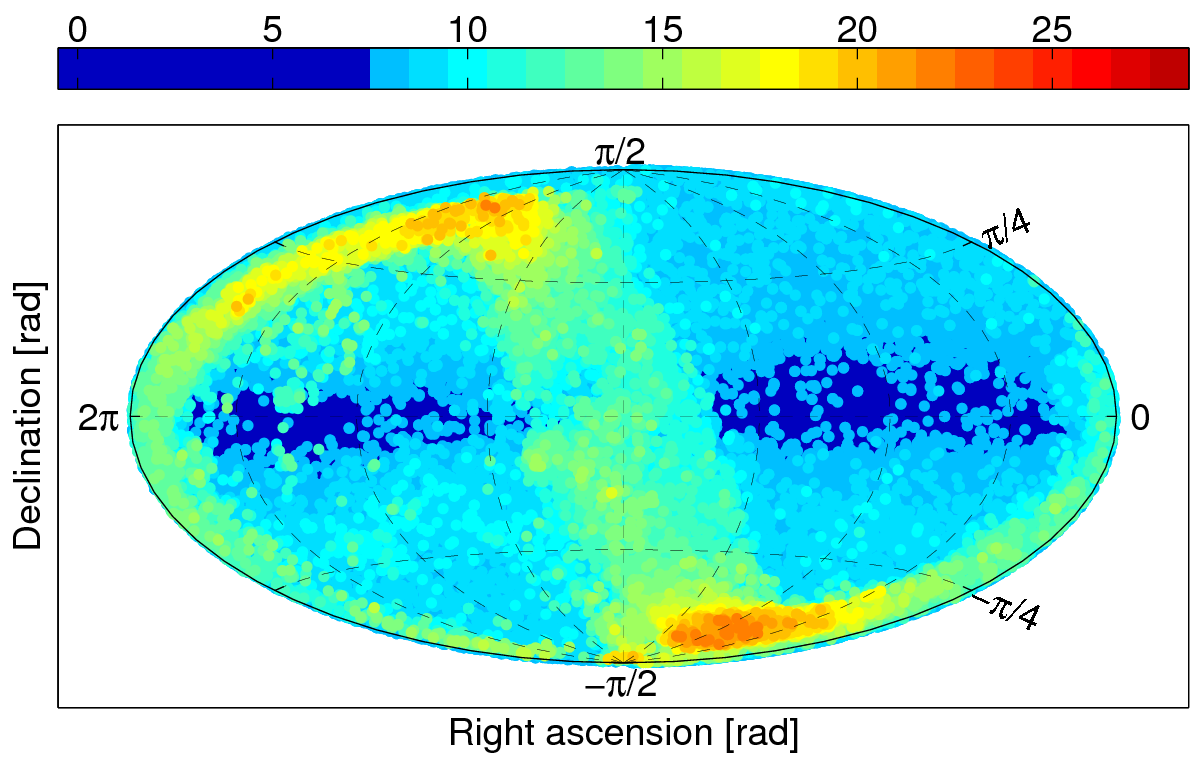}}\\
	\subfigure[\label{f:skym-c}]
	{\includegraphics[width=0.85\columnwidth]{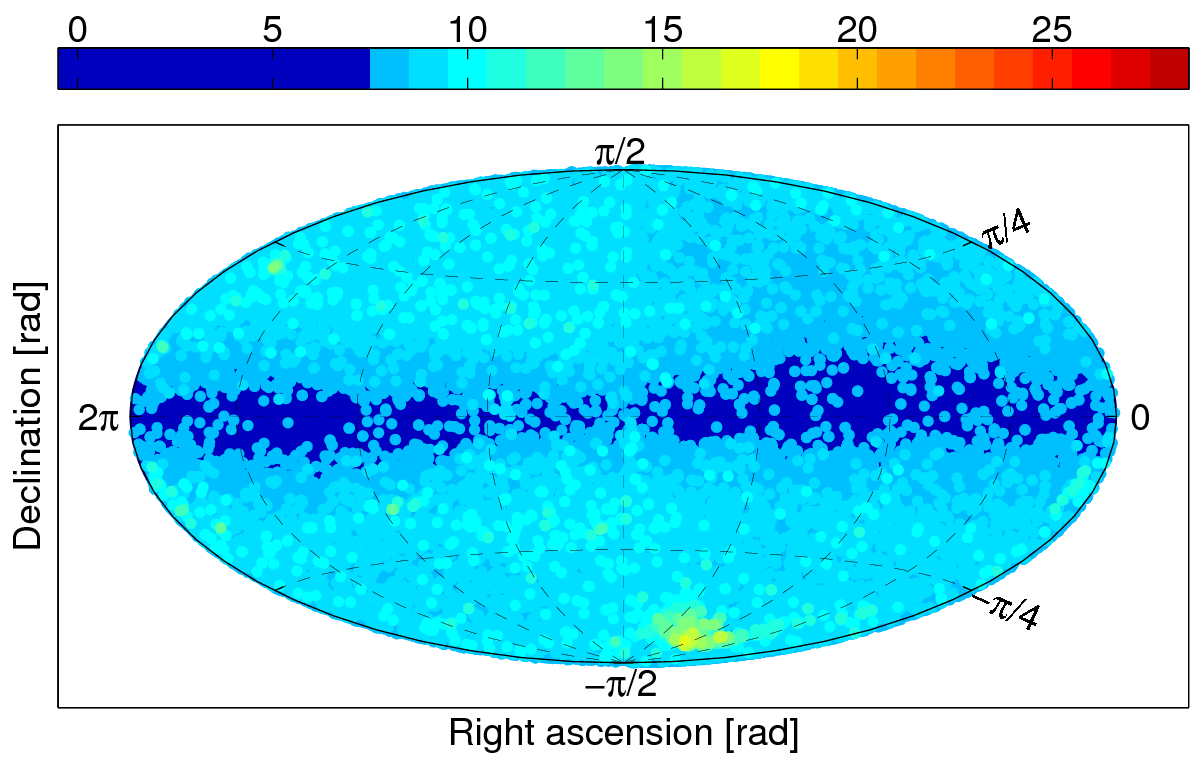}}\\
	\subfigure[\label{f:skym-d}]
	{\includegraphics[width=0.85\columnwidth]{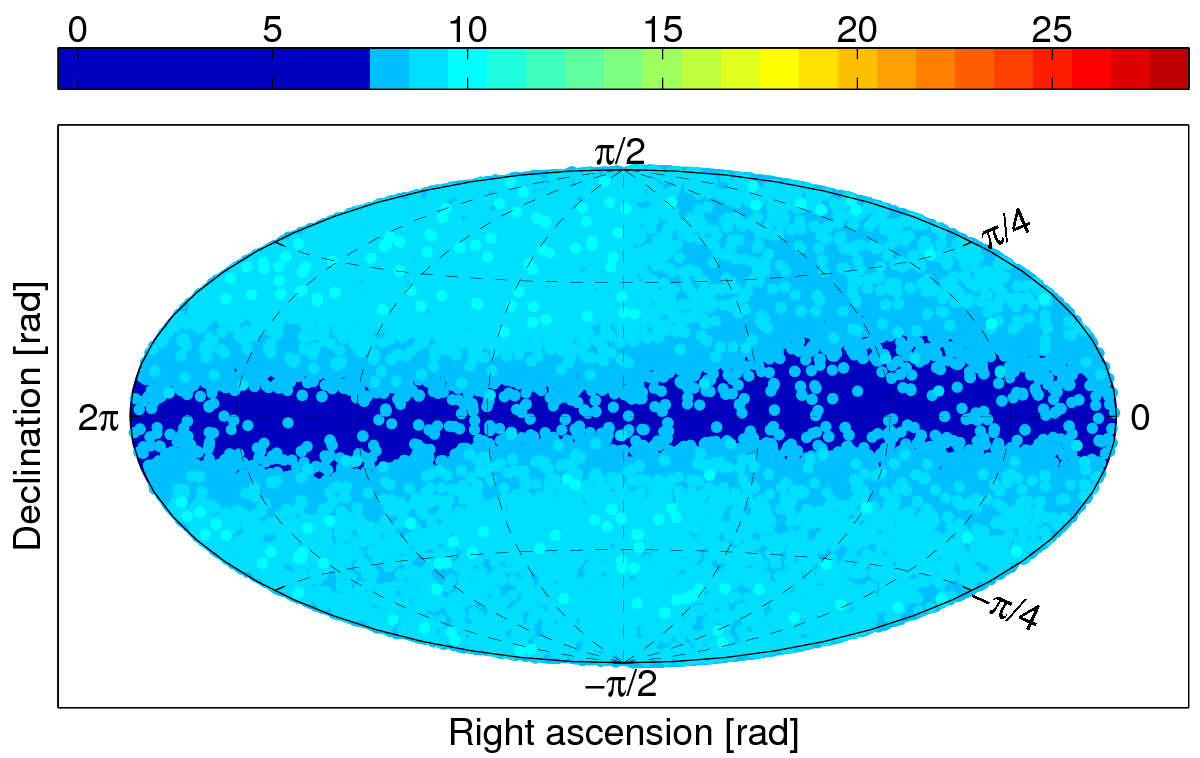}}
         \caption{ Sky maps of post-processing results. 
          Candidates having more than $7$ coincidences are shown in
          Hammer-Aitoff projections of the sky.  The color-bar
          indicates the number of coincidences of a particular candidate (cell).  
          The top plot~\subref{f:skym-a} shows the coincidence analysis results.
          In~\subref{f:skym-c}, {\it a posteriori}  strong lines of known instrumental 
          origin and hardware injections are removed.
          The bottom plot~\subref{f:skym-d} is obtained by additionally applying 
          the parameter-space veto and excluding
	 single-detector candidates.
          \label{f:AllResults} }
\end{figure}

\begin{figure}
         \subfigure[\label{f:spec-a}]
         {\includegraphics[width=1.0\columnwidth]{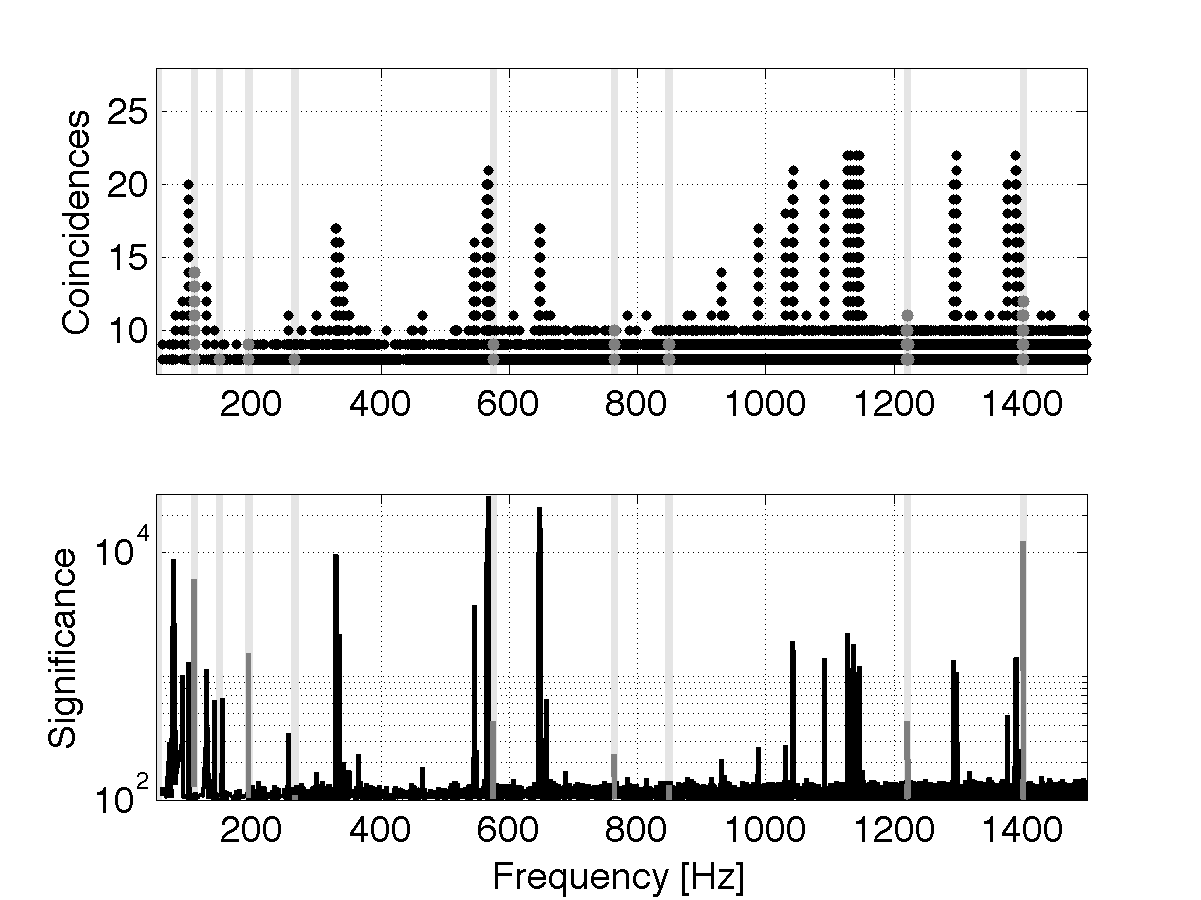}}
         \subfigure[\label{f:spec-b}]
         {\includegraphics[width=1.0\columnwidth]{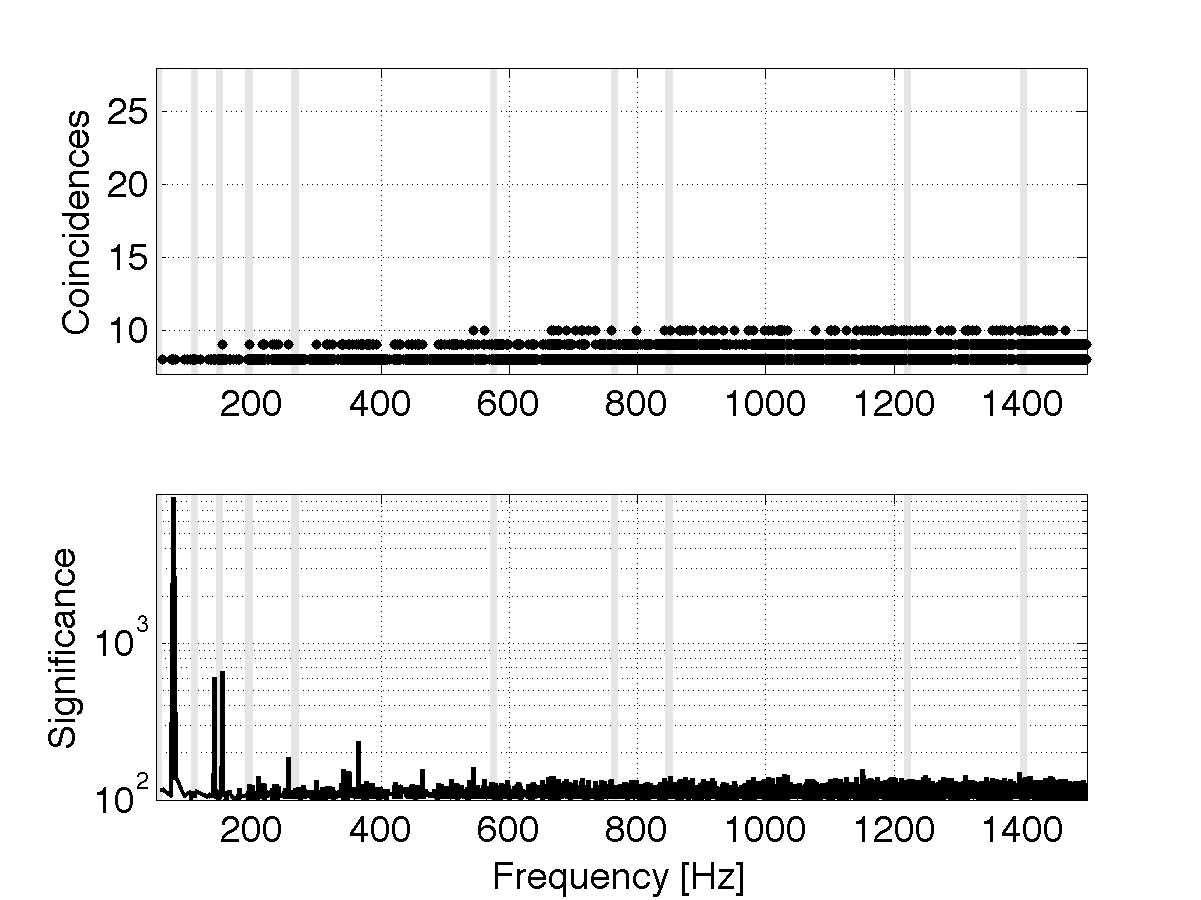}}
         \caption{
          The top plot~\subref{f:spec-a} shows the post-processing candidates having 
          more than $7$ coincidences as function of frequency.
          The light-gray shaded rectangular regions highlight the frequency bands of
          the hardware injections.
          The dark-gray data points show the candidates resulting from the
          hardware-injected GW signals. 
          In~\subref{f:spec-b}, the final results are shown after exclusion of 
          instrumental lines of known origin and hardware injections,
          application of parameter-space veto and exclusion of single-detector 
          candidates.
          \label{f:specplot}
        }
 \end{figure}

Figures~\ref{f:AllResults} and~\ref{f:specplot}  summarize
all post-processing results from the entire search frequency range of
$50\,\Hz$ to $1500\,\Hz$, for each frequency coincidence cell
maximized over the entire sky and full spin-down range.


In Fig.~\ref{f:skym-a} all candidates that have \mbox{$7$ or more} 
coincidences are shown in a sky projection.
The color scale is used to indicate the number of coincidences.
The most prominent feature still apparent forms an annulus of 
high coincidences in the sky, including the ecliptic poles, a distinctive 
fingerprint of the instrumental noise lines~\cite{PletschGC}.
To obtain the results shown in Fig.~\ref{f:skym-c},
the set of candidates is cleaned {\it a posteriori} by removing strong 
instrumental noise lines, whose origin became understood after
the search was begun, and excluding
the  hardware injections. 
Finally, in Fig.~\ref{f:skym-d} the parameter-space
veto is applied and coincidence cells which contain candidate events from
a single detector only are excluded, too.

In Fig.~\ref{f:spec-a} the coincidences and significance of all
candidates that have \mbox{$7$ or more} coincidences are shown as a function
of frequency.  From this set of candidates the hardware injections are excluded,
strong instrumental noise lines of known origin are removed, the parameter-space
veto is applied and finally single-detector candidates are excluded
to obtain Fig.~\ref{f:spec-b}.

As can be seen from Figs.~\ref{f:skym-d} and~\ref{f:spec-b} there are
no candidates that exceed the predefined detection threshold of
$20$ coincidences (which would initiate more a extensive investigation).
The largest number of coincidences found is~$10$, which is at the
background level of false alarms expected from
random noise only.
From these candidates having $10$ coincidences, 
Table~\ref{t:topten} lists the ten most significant ones.

\begin{table*}[htb!p]
\caption{The ten most significant post-processing candidates that have $10$ or more coincidences.
  The frequency of each candidate~$f_{\rm cand}$ refers to the fiducial GPS time
  \mbox{$t_{\rm{fiducial}} = 816\,397\,490\,{\rm s}$}.
  The parameters
  $\delta_{\rm cand}$, $\alpha_{\rm cand}$, $\dot f_{\rm cand}$,
  $\mathcal{C}_{\rm cand} = \mathcal{C}^{\rm H1}_{\rm cand} +
  \mathcal{C}^{\rm L1}_{\rm cand}$ and  $\mathcal{S}_{\rm cand}$ are for the
  most-significant, most-coincident candidate with the given frequency
  of $f_{\rm cand}$, where
  $ \mathcal{C}^{\rm H1}_{\rm cand}$ and  $\mathcal{C}^{\rm L1}_{\rm cand}$
  denote the number of coincidences 
  from detectors H1 and L1, respectively.
  \label{t:topten}}
\begin{tabular}{crrrrccccrr} \hline
&$f_{\rm cand}$ [Hz] &$\delta_{\rm cand}$ [rad] & $\alpha_{\rm cand}$ [rad] & $\dot f_{\rm cand}$ [$\mathrm{Hz}~\mathrm{s}^{-1}$] & $\mathcal{C}_{\rm cand}$ & $\mathcal{C}^{\rm H1}_{\rm cand}$ & $\mathcal{C}^{\rm L1}_{\rm cand}$ & \; $\mathcal{S}_{\rm cand}$ \;& $P_{\rm F}$ per 0.5Hz\\
\hline \hline
 & $543.810438$ & $0.6823$ & $5.9944$ & $-3.24\times 10^{-10}$ & $10$ & $8$ & $2$ & $160.9$ & $7.2\times 10^{-5}$ &\\
 & $1151.534608$ & $1.1330$ & $5.4462$ & $2.11\times 10^{-11}$ & $10$ & $4$ & $6$ & $154.3$ & $1.4\times 10^{-3}$ &\\
 & $1395.351068$ & $-1.1928$ & $2.5980$ & $-3.92\times 10^{-9}$ & $10$ & $8$ & $2$ & $150.4$ & $7.1\times 10^{-4}$ &\\
 & $1249.855062$ & $-1.2380$ & $6.0203$ & $-2.43\times 10^{-9}$ & $10$ & $8$ & $2$ & $144.2$ & $4.5\times 10^{-3}$ &\\
 & $1311.458030$ & $-0.5143$ & $6.1638$ & $-3.32\times 10^{-9}$ & $10$ & $8$ & $2$ & $142.8$ & $1.7\times 10^{-3}$ &\\
 & $1033.967720$ & $0.6002$ & $5.3133$ & $-1.83\times 10^{-9}$ & $10$ & $8$ & $2$ & $142.7$ & $1.2\times 10^{-3}$ &\\
 & $851.799376$ & $1.1071$ & $3.2019$ & $-7.79\times 10^{-10}$ & $10$ & $8$ & $2$ & $142.1$ & $4.1\times 10^{-4}$ &\\
 & $665.944644$ & $-0.4602$ & $2.3638$ & $-1.28\times 10^{-9}$ & $10$ & $6$ & $4$ & $141.9$ & $1.0\times 10^{-3}$ &\\
 & $669.187638$ & $-0.6928$ & $3.0333$ & $-1.58\times 10^{-9}$ & $10$ & $7$ & $3$ & $141.6$ & $1.0\times 10^{-3}$ &\\
 & $1443.831722$ & $0.7046$ & $6.0788$ & $-4.47\times 10^{-9}$ & $10$ & $7$ & $3$ & $141.5$ & $3.5\times 10^{-3}$ &\\
\hline
\end{tabular}
\end{table*}

\section{Comparison with previous searches\label{sec:Comparison}}

A previous paper~\cite{S4EAH} reported on the results of the 
Einstein@Home search for periodic GW signals in the LIGO S4 data.
The present work extends this search analyzing more sensitive LIGO S5 data
while using the same methods described in~\cite{S4EAH}. Therefore,
this section elucidates the changes in configuration of the search and
post-processing.

First, not only is more sensitive data used here, but a larger total
volume of data is searched compared to~\cite{S4EAH}. The number
of $30$-h data segments analyzed increased from $17$ to $28$. 

In addition, the template grids used in each data segment of this search
were constructed to be denser, reducing the possible loss
of signals due to mismatch in the template waveforms.
Compared to the previous search in S4 data, where a maximal mismatch of $m=0.2$ ($m=0.5$) was
used in the low (high) frequency range, here templates are placed on a
grid of higher density using $m=0.15$ ($m=0.4$)
in the low (high) frequency range.

Moreover, in the high-frequency range a larger
range of possible spin-downs is searched. The S4 analysis searched over minimum
spin-down ages greater than $10\,000\,{\rm yr}$ for frequencies
in the higher range ($f> 300\,\Hz$), whereas this analysis
searches over minimum
spin-down ages greater than $8\,000\,{\rm yr}$ for frequencies
in the higher range ($f> 400\,\Hz$). The different partitioning
of frequencies into the low and high ranges (split at $300\,\Hz$
in S4, split at $400\,\Hz$ here) is a consequence of an
optimization study reflecting the overall most sensitive search
at given computing power.

This search presented here analyzed in total 
about three times more workunits than in the S4 search.
In searching the S4 data, each workunit returned the top $13\,000$ candidate
events, whereas this search is designed to keep only the 
top $1\,000$ ($10\,000$) candidate events in the low (high) frequency 
range. This configuration has the purpose of balancing the load
on the Einstein@Home servers, which receive the workunit results.
A low-frequency workunit returns a factor of $10$ fewer events, because these were 
designed to last approximately $10$ times less than each high-frequency workunit.

Finally, based on the estimates presented in Sec.~\ref{sec:ExpectSen}, 
the present search is overall about a factor of three more sensitive 
than the previous S4 search. This improvement is a consequence of using more
sensitive detector data in combination with a finer-spaced 
template bank.

\begin{figure}
   \includegraphics[width=1.0\columnwidth]{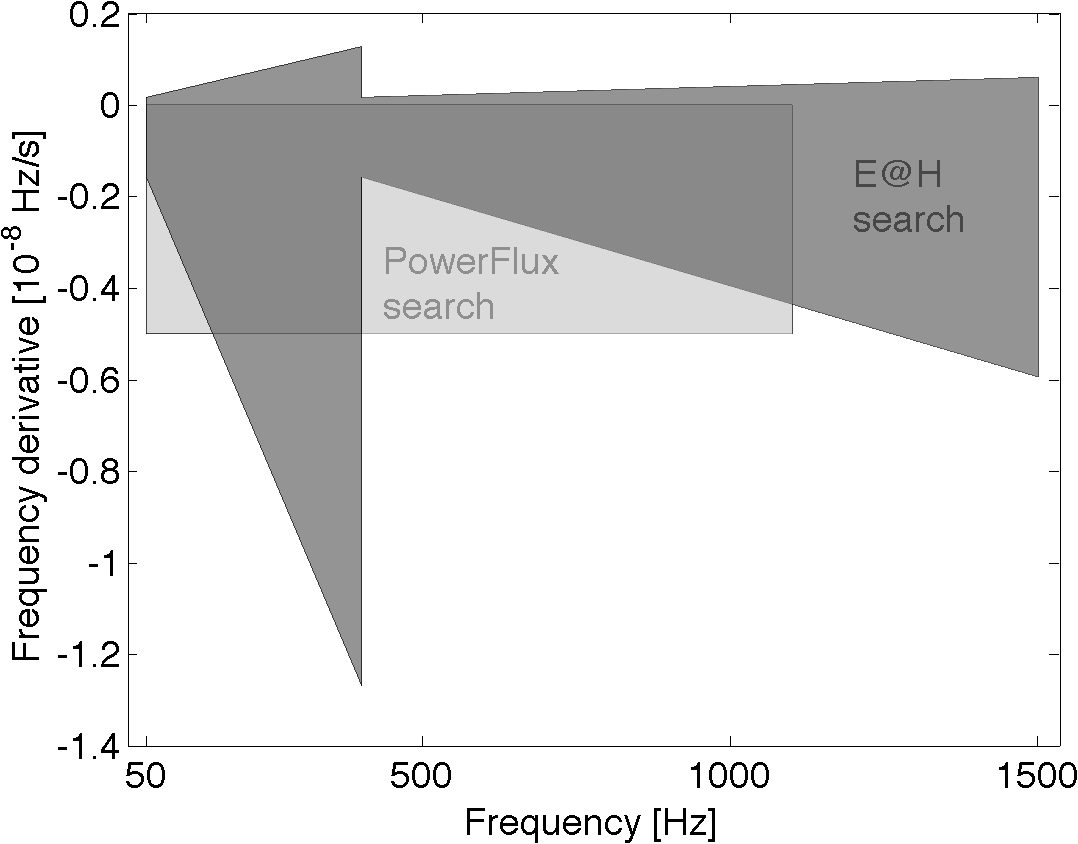}
   \caption{Comparison of search parameter spaces in the plane of frequency and frequency derivative.
   The dark-gray region refers to this Einstein@Home all-sky analysis in early LIGO S5 data.
   The light-grey area corresponds to the recent all-sky PowerFlux search~\cite{PFearlyS5} in early LIGO S5 data.}
   \label{f:searchsp}
\end{figure}

The methods used here, as well as in the S4 paper, would be expected
to give very high confidence if a strong enough signal were present in the data.
It is interesting to compare the sensitivity of this detection scheme with
the sensitivity of upper limits such as presented recently in~\cite{PFearlyS5}.
Based on the PowerFlux method~\cite{S4IncoherentPaper}, that analysis set
strain upper limits at the $95\%$ confidence level in the frequency range 
of \mbox{50 -- 1100 Hz} and the frequency-derivative range 
of \mbox{$-5\times10^{-9}-0\,{\rm Hz\, s^{-1}}$} using $7\,147\,{\rm h}$ of early LIGO S5 data,
about $8.5$ times more data than was used here.
Note that this Einstein@Home search explores substantially larger parts 
of parameter space in frequency and frequency derivative, as shown in Fig.~\ref{f:searchsp}.

The upper-limit worst-case results of~\cite{PFearlyS5} for the equatorial
sky region are remarkably close to the $90\%$-detection-level $h_0$-values 
of Fig.~\ref{f:estULs}. However, these PowerFlux upper limits refer to the most 
unfavorable polarization and sky position. A population-based upper limit
over all sky locations and polarizations would be lower.

On the other hand, another key difference between the PowerFlux upper limits 
procedure and the sensitivity estimation carried out here is the detection criteria.
In the present work, detection requires a signal to generate $20$ or more
coincidences among the $28$ different data segments. This corresponds
to a false alarm probability in Gaussian noise of the order $10^{-21}$ per
$0.5\,{\rm Hz}$ frequency band. This is different from~\cite{PFearlyS5},
where simulated signals are compared to the strongest candidates found.
Thus, an equivalent detection criterion for this work would be to compare
the signals against the strongest candidates in each $0.5\,{\rm Hz}$ band.
These are typically $10$ coincidences, which relates to a Gaussian noise
false alarm rate of order $10^{-3}$. One can estimate the effect on sensitivity
by recomputing the sensitivity estimation of Sec.~\ref{sec:ExpectSen}, but
requiring each signal to produce only $10$ coincidences. This reduces
the prefactors $R_{\D}$ given above by a factor of $1.24$. 

Apart from the larger parameter space searched,
the present analysis is achieving roughly comparable sensitivity 
to \cite{PFearlyS5} in spite of searching $8.5$ times less data.
Much of this effectiveness is due to the increased coherent integration 
time ($30$ hours vs $30$ minutes), which is only possible due to the great amount 
of computing power donated by the tens of thousands of Einstein@Home volunteers.

\section{Conclusion \label{sec:Conclusion}}

Using early LIGO fifth-science-run data this paper reports on the results from 
the Einstein@Home search for unknown periodic GW sources,
extending the previous Einstein@Home search in LIGO S4 data~\cite{S4EAH}. 
The sensitivity of the present analysis improves upon the
previous Einstein@Home S4 search by a factor of about three.
Additionally, in large regions of the parameter space probed, this analysis yields
the currently most sensitive all-sky search results for periodic GW sources.

No credible periodic GW signal was found. Over a $100$-Hz wide band
around the detectors' most sensitive frequencies, more than
$90\%$ of sources with dimensionless gravitational-wave strain amplitude
greater than $3\times10^{-24}$ would have been detected.

While no statistically significant signal was observed in this analysis, 
the results demonstrate the capability of public distributed
computing to accomplish a sensitive periodic GW search for 
the benefit of future searches.

The sensitivity of the present analysis is essentially limited
by the first-stage threshold on $\F$-statistics forced by
the limited data volume which can be returned from the participating clients.
A new Einstein@Home search currently underway carries out 
the incoherent combination of $\F$-statistic results 
on the client machines (done here in the post-processing once results were sent back).
This makes it possible to set a much lower (sensitivity-optimized) first-stage threshold 
on $\F$-statistics. Hence, results from the new search promise a
significant enhancement in the overall sensitivity for a periodic
GW detection.

\section*{Acknowledgments\label{sec:Ack}}
\input acknowledgements.tex
This document has been assigned LIGO Laboratory document number~\dcc.

\begin{appendix}

\section{Cleaned instrumental noise lines
\label{sec:aprioricleaning}}

Table~\ref{t:lines} lists the frequencies of noise lines excluded
from the data and replaced by Gaussian noise {\it a priori} to the search. 
\begin{table*}
\caption{Instrumental noise lines cleaned from H1 and L1 data.
The three columns show the central frequency~$f_{\rm{Line}}$,
the bandwidth~$\Delta f_{\rm{Line}}^{(1)}$ removed below the
central frequency and the bandwidth~$\Delta f_{\rm{Line}}^{(2)}$
removed above the central frequency. Thus the total bandwidth
removed per central frequency is $\Delta f_{\rm{Line}}^{(1)}+\Delta f_{\rm{Line}}^{(2)}$.
In addition, at {\it each harmonic} of the $60\,\Hz$ mains frequency, 
the same bandwidth is also removed.
A zero bandwidth indicates that the line-cleaning algorithm replaces in these cases 
a single Fourier bin with the average of bins on either side.  
The spacing between Fourier bins is $1/1800\,\Hz$.
\label{t:lines}}
\begin{center}
\begin{minipage}[b]{0.55\columnwidth}
\begin{tabular}[t]{rcc}
\hline
& H1  & \\
$f_{\rm{Line}}$[Hz] & $\Delta f_{\rm{Line}}^{(1)}$[Hz]& $\Delta f_{\rm{Line}}^{(2)}$[Hz] \\
\hline\hline
46.7 & 0.0 & 0.0 \\
60.0 & 1.0 & 1.0 \\
346.0 & 4.0 & 4.0 \\
393.1 & 0.0 & 0.0 \\
686.9 & 0.3 & 0.3 \\
688.2 & 0.3 & 0.3 \\
689.5 & 0.5 & 0.6 \\
694.75 & 1.25 & 1.25 \\
1030.55 & 0.1 & 0.1 \\
1032.18 & 0.04 & 0.04 \\
1032.58 & 0.1 & 0.1 \\
1033.7 & 0.1 & 0.1 \\
1033.855 & 0.05 & 0.05 \\
1034.6 & 0.4 & 0.4 \\
1041.23 & 0.1 & 0.1 \\
1042.0 & 0.5 & 0.2 \\
1043.4 & 0.2 & 0.2 \\
1144.3 & 0.0 & 0.0 \\
1373.75 & 0.1 & 0.1 \\
1374.44 & 0.1 & 0.1 \\
1377.14 & 0.1 & 0.1 \\
1378.75 & 0.1 & 0.1 \\
1379.52 & 0.1 & 0.1 \\
1389.06 & 0.06 & 0.06 \\
1389.82 & 0.07 & 0.07 \\
1391.5 & 0.2 & 0.2 \\
\hline
\end{tabular}
\end{minipage}
\qquad
\begin{minipage}[t]{0.55\columnwidth}
\begin{tabular}[t]{rcc}
\hline
& L1  & \\
$f_{\rm{Line}}$[Hz] & $\Delta f_{\rm{Line}}^{(1)}$[Hz]& $\Delta f_{\rm{Line}}^{(2)}$[Hz] \\
\hline\hline
54.7 & 0.0 & 0.0 \\
60.0 & 1.0 & 1.0 \\
345.0 & 5.0 & 5.0 \\
396.7 & 0.0 & 0.0 \\
686.5 & 1.0 & 1.0 \\
688.83 & 0.5 & 0.5 \\
693.7 & 0.7 & 0.7 \\
1029.5 & 0.25 & 0.25 \\
1031 & 0.5 & 0.5 \\
1033.6 & 0.2 & 0.2 \\
1041 & 1.0 & 1.0 \\
1151.5 & 0.0 & 0.0 \\
1372.925 & 0.075 & 0.075 \\
1374.7 & 0.1 & 0.1 \\
1375.2 & 0.1 & 0.1 \\
1378.39 & 0.1 & 0.1 \\
1387.4 & 0.05 & 0.05 \\
1388.5 & 0.3 & 0.3 \\
\hline
\end{tabular}
\end{minipage}
\end{center}
\end{table*}
Table~\ref{t:postcleaning} lists the central frequencies around either side 
of which the Doppler band \mbox{($\Delta f_{\rm{Line}}=f_{\rm{Line}}\times10^{-4}$)} 
is {\it a posteriori} excluded from the post-processed search results.
\begin{table}[htb!p]
\caption{Frequencies of instrumental lines that have been 
excluded {\it a posteriori} from the post-processed
search results. Each column shows the central frequency~$f_{\rm{Line}}$
around which a bandwidth of \mbox{$\Delta f_{\rm{Line}}=f_{\rm{Line}}\times10^{-4}$} has been removed
on either side. The cleaned bandwidth corresponds to
the maximum possible frequency shift due to the global parameter-space
correlations~\cite{PletschGC}. On physical grounds this is related to
the maximum possible Doppler shift due to the orbital velocity 
of the Earth, which is approximately  $10^{-4}$ in units of the speed of light.
\\ \label{t:postcleaning}}
\begin{minipage}{0.23\columnwidth}
\begin{tabular}{crc}
\hline
&$f_{\rm{Line}}$[Hz] & \\
\hline\hline
& 69.75 &\\
& 90.0   &\\
& 100.0 &\\
& 128.0 &\\
& 256.0 &\\
& 335.0 &\\
& 329.0 &\\
& 546.01 &\\
& 548.38 &\\
& 564.14 &\\
& 566.17 &\\
\hline
\end{tabular}
\end{minipage}
\begin{minipage}{0.23\columnwidth}
\begin{tabular}{crc}
\hline
&$f_{\rm{Line}}$[Hz] & \\
\hline\hline
& 568.17 &\\
& 570.41 &\\
& 645.56 &\\
& 646.46 &\\
& 647.07 &\\
& 648.84 &\\
& 649.46 &\\
& 658.74 &\\
& 686.92 &\\
& 930.34 &\\
& 988.19 &\\
\hline
\end{tabular}
\end{minipage}
\begin{minipage}{0.23\columnwidth}
\begin{tabular}{crc}
\hline
&$f_{\rm{Line}}$[Hz] & \\
\hline\hline
& 1030.55 &\\
& 1042.19 &\\
& 1043.33 &\\
& 1092.01 &\\
& 1128.28 &\\
& 1132.22 &\\
& 1136.23 &\\
& 1142.87 &\\
& 1145.29 &\\
& 1146.59 &\\
& 1291.11 &\\
\hline
\end{tabular}
\end{minipage}
\begin{minipage}{0.23\columnwidth}
\begin{tabular}{crc}
\hline
&$f_{\rm{Line}}$[Hz] & \\
\hline\hline
& 1292.91 &\\
& 1294.14 &\\
& 1297.67 &\\
& 1298.93 &\\
& 1317.47 &\\
& 1377.14 &\\
& 1388.38 &\\
& 1390.70 &\\
& 1391.60 &\\
&&\\
&&\\
\hline
\end{tabular}
\end{minipage}
\end{table}

\end{appendix}

\bibliography{S5R1_EAH}

\end{document}

%% file: authorlist.tex
%********************* cut here mark*******************
%NOTE ADD altafillletter to \documentclass if lettered affiliation footnotes are desired
% EXAMPLE:
%
%\documentclass[prd,superscriptaddress,showpacs,amssymb,amsmath,amsfonts,aps,altaffilletter]{revtex4}
%
%%%%%%%%%%%% Institutes Number and definitions %%%%%%%%%%%
%
% This list was generated on 4 Feb 2009 A4 DCC LIGO - T0900008-v4

\newcommand*{\AG}{Albert-Einstein-Institut, Max-Planck-Institut f\"{u}r Gravitationsphysik, D-14476 Golm, Germany}
\affiliation{\AG}
\newcommand*{\AH}{Albert-Einstein-Institut, Max-Planck-Institut f\"{u}r Gravitationsphysik, D-30167 Hannover, Germany}
\affiliation{\AH}
\newcommand*{\AU}{Andrews University, Berrien Springs, MI 49104 USA}
\affiliation{\AU}
\newcommand*{\AN}{Australian National University, Canberra, 0200, Australia}
\affiliation{\AN}
\newcommand*{\CH}{California Institute of Technology, Pasadena, CA  91125, USA}
\affiliation{\CH}
\newcommand*{\CA}{Caltech-CaRT, Pasadena, CA  91125, USA}
\affiliation{\CA}
\newcommand*{\CU}{Cardiff University, Cardiff, CF24 3AA, United Kingdom}
\affiliation{\CU}
\newcommand*{\CL}{Carleton College, Northfield, MN  55057, USA}
\affiliation{\CL}
\newcommand*{\CS}{Charles Sturt University, Wagga Wagga, NSW 2678, Australia}
\affiliation{\CS}
\newcommand*{\CO}{Columbia University, New York, NY  10027, USA}
\affiliation{\CO}
\newcommand*{\ER}{Embry-Riddle Aeronautical University, Prescott, AZ   86301 USA}
\affiliation{\ER}
\newcommand*{\EU}{E\"{o}tv\"{o}s University, ELTE 1053 Budapest, Hungary}
\affiliation{\EU}
\newcommand*{\HC}{Hobart and William Smith Colleges, Geneva, NY  14456, USA}
\affiliation{\HC}
\newcommand*{\IA}{Institute of Applied Physics, Nizhny Novgorod, 603950, Russia}
\affiliation{\IA}
\newcommand*{\IU}{Inter-University Centre for Astronomy  and Astrophysics, Pune - 411007, India}
\affiliation{\IU}
\newcommand*{\HU}{Leibniz Universit\"{a}t Hannover, D-30167 Hannover, Germany}
\affiliation{\HU}
\newcommand*{\CT}{LIGO - California Institute of Technology, Pasadena, CA  91125, USA}
\affiliation{\CT}
\newcommand*{\LO}{LIGO - Hanford Observatory, Richland, WA  99352, USA}
\affiliation{\LO}
\newcommand*{\LV}{LIGO - Livingston Observatory, Livingston, LA  70754, USA}
\affiliation{\LV}
\newcommand*{\LM}{LIGO - Massachusetts Institute of Technology, Cambridge, MA 02139, USA}
\affiliation{\LM}
\newcommand*{\LU}{Louisiana State University, Baton Rouge, LA  70803, USA}
\affiliation{\LU}
\newcommand*{\LE}{Louisiana Tech University, Ruston, LA  71272, USA}
\affiliation{\LE}
\newcommand*{\LL}{Loyola University, New Orleans, LA 70118, USA}
\affiliation{\LL}
\newcommand*{\MT}{Montana State University, Bozeman, MT 59717, USA}
\affiliation{\MT}
\newcommand*{\MS}{Moscow State University, Moscow, 119992, Russia}
\affiliation{\MS}
\newcommand*{\ND}{NASA/Goddard Space Flight Center, Greenbelt, MD  20771, USA}
\affiliation{\ND}
\newcommand*{\NA}{National Astronomical Observatory of Japan, Tokyo  181-8588, Japan}
\affiliation{\NA}
\newcommand*{\NO}{Northwestern University, Evanston, IL  60208, USA}
\affiliation{\NO}
\newcommand*{\RI}{Rochester Institute of Technology, Rochester, NY  14623, USA}
\affiliation{\RI}
\newcommand*{\RA}{Rutherford Appleton Laboratory, HSIC, Chilton, Didcot, Oxon OX11 0QX United Kingdom}
\affiliation{\RA}
\newcommand*{\SJ}{San Jose State University, San Jose, CA 95192, USA}
\affiliation{\SJ}
\newcommand*{\SM}{Sonoma State University, Rohnert Park, CA 94928, USA}
\affiliation{\SM}
\newcommand*{\SE}{Southeastern Louisiana University, Hammond, LA  70402, USA}
\affiliation{\SE}
\newcommand*{\SO}{Southern University and A\&M College, Baton Rouge, LA  70813, USA}
\affiliation{\SO}
\newcommand*{\SA}{Stanford University, Stanford, CA  94305, USA}
\affiliation{\SA}
\newcommand*{\SR}{Syracuse University, Syracuse, NY  13244, USA}
\affiliation{\SR}
\newcommand*{\PU}{The Pennsylvania State University, University Park, PA  16802, USA}
\affiliation{\PU}
\newcommand*{\UM}{The University of Melbourne, Parkville VIC 3010, Australia}
\affiliation{\UM}
\newcommand*{\MI}{The University of Mississippi, University, MS 38677, USA}
\affiliation{\MI}
\newcommand*{\SF}{The University of Sheffield, Sheffield S10 2TN, United Kingdom}
\affiliation{\SF}
\newcommand*{\TA}{The University of Texas at Austin, Austin, TX 78712, USA}
\affiliation{\TA}
\newcommand*{\TC}{The University of Texas at Brownsville and Texas Southmost College, Brownsville, TX  78520, USA}
\affiliation{\TC}
\newcommand*{\TR}{Trinity University, San Antonio, TX  78212, USA}
\affiliation{\TR}
\newcommand*{\BB}{Universitat de les Illes Balears, E-07122 Palma de Mallorca, Spain}
\affiliation{\BB}
\newcommand*{\UA}{University of Adelaide, Adelaide, SA 5005, Australia}
\affiliation{\UA}
\newcommand*{\BR}{University of Birmingham, Birmingham, B15 2TT, United Kingdom}
\affiliation{\BR}
\newcommand*{\BE}{University of California at Berkeley, Berkeley, CA 94720 USA}
\affiliation{\BE}
\newcommand*{\FA}{University of Florida, Gainesville, FL  32611, USA}
\affiliation{\FA}
\newcommand*{\GU}{University of Glasgow, Glasgow, G12 8QQ, United Kingdom}
\affiliation{\GU}
\newcommand*{\MD}{University of Maryland, College Park, MD 20742 USA}
\affiliation{\MD}
\newcommand*{\AM}{University of Massachusetts - Amherst, Amherst, MA 01003, USA}
\affiliation{\AM}
\newcommand*{\MU}{University of Michigan, Ann Arbor, MI  48109, USA}
\affiliation{\MU}
\newcommand*{\MN}{University of Minnesota, Minneapolis, MN 55455, USA}
\affiliation{\MN}
\newcommand*{\OU}{University of Oregon, Eugene, OR  97403, USA}
\affiliation{\OU}
\newcommand*{\RO}{University of Rochester, Rochester, NY  14627, USA}
\affiliation{\RO}
\newcommand*{\SL}{University of Salerno, 84084 Fisciano (Salerno), Italy}
\affiliation{\SL}
\newcommand*{\SN}{University of Sannio at Benevento, I-82100 Benevento, Italy}
\affiliation{\SN}
\newcommand*{\SH}{University of Southampton, Southampton, SO17 1BJ, United Kingdom}
\affiliation{\SH}
\newcommand*{\SC}{University of Strathclyde, Glasgow, G1 1XQ, United Kingdom}
\affiliation{\SC}
\newcommand*{\WA}{University of Western Australia, Crawley, WA 6009, Australia}
\affiliation{\WA}
\newcommand*{\UW}{University of Wisconsin-Milwaukee, Milwaukee, WI  53201, USA}
\affiliation{\UW}
\newcommand*{\WU}{Washington State University, Pullman, WA 99164, USA}
\affiliation{\WU}

\author{}    \affiliation{\GU}    
\author{B.~P.~Abbott}    \affiliation{\CT}    
\author{R.~Abbott}    \affiliation{\CT}    
\author{R.~Adhikari}    \affiliation{\CT}    
\author{P.~Ajith}    \affiliation{\AH}    
\author{B.~Allen}    \affiliation{\AH}  \affiliation{\UW}  
\author{G.~Allen}    \affiliation{\SA}    
\author{R.~S.~Amin}    \affiliation{\LU} 
\author{S.~B.~Anderson}    \affiliation{\CT}    
\author{W.~G.~Anderson}    \affiliation{\UW}    
% ...
 \collaboration{The LIGO Scientific Collaboration, http://www.ligo.org}
 \noaffiliation
 \author{D.~P.~Anderson} \affiliation{\BE}   
%
%
%********************* cut here mark*******************

%% file: acknowledgements.tex
The authors thank the tens of thousands of volunteers who have supported
the Einstein@Home project by donating their computer time and expertise for this
analysis. Without their contributions, this work would not have been possible.

The authors gratefully acknowledge the support of the United States
National Science Foundation for the construction and operation of the
LIGO Laboratory and the Science and Technology Facilities Council of the
United Kingdom, the Max-Planck-Society, and the State of
Niedersachsen/Germany for support of the construction and operation of
the GEO600 detector. The authors also gratefully acknowledge the support
of the research by these agencies and by the Australian Research Council,
the Council of Scientific and Industrial Research of India, the Istituto
Nazionale di Fisica Nucleare of Italy, the Spanish Ministerio de
Educaci\'on y Ciencia, the Conselleria d'Economia, Hisenda i Innovaci\'o of
the Govern de les Illes Balears, the Royal Society, the Scottish Funding 
Council, the Scottish Universities Physics Alliance, The National Aeronautics 
and Space Administration, the Carnegie Trust, the Leverhulme Trust, the David
and Lucile Packard Foundation, the Research Corporation, and the Alfred
P. Sloan Foundation.